\documentclass[aps,prc,twocolumn,letterpaper,floatfix]{revtex4}
\usepackage{graphicx,amsmath,amssymb}
\usepackage[usenames,dvipsnames]{pstricks}
\usepackage{pst-grad}
\usepackage{url}
\usepackage{xfrac}

\newcommand{\ket}[1]{| {#1} \rangle}
\newcommand{\bra}[1]{\langle {#1} |}
\newcommand{\bracket}[2]{\langle {#1} | {#2} \rangle}

\begin{document}

\title{Weak Measurement and (Bohmian) Conditional Wave Functions}
\author{Travis Norsen}
\affiliation{Physics Department, Smith College, Northampton, MA 01060, USA.}
\email{travisnorsen@gmail.com}

\author{Ward Struyve}
\affiliation{Departments of Mathematics and Philosophy, Rutgers University, Hill Center, 110 Frelinghuysen Road, Piscataway, NJ 08854-8019, USA.}
\email{wstruyve@math.rutgers.edu}

\date{\today}

\begin{abstract}
It was recently pointed out (and demonstrated experimentally) 
by Lundeen \emph{et al.}\  that 
the wave function of a particle
(more precisely, the wave function possessed by each member
of an ensemble of identically-prepared particles) can be
``directly measured'' using weak measurement.  Here it is
shown that if this same technique is applied, with appropriate
post-selection, to one particle from a
(perhaps entangled) multi-particle system, the result is precisely the
so-called ``conditional wave function'' of Bohmian mechanics.  Thus, a
plausibly operationalist method for defining the wave function of a
quantum mechanical sub-system corresponds to the natural definition of
a sub-system wave function which Bohmian mechanics (uniquely) makes
possible.  Similarly, a weak-measurement-based procedure for
directly measuring a sub-system's density matrix should yield, under
appropriate circumstances, the Bohmian ``conditional density matrix'' as
opposed to the standard reduced density matrix.  Experimental
arrangements to demonstrate this behavior -- and also thereby reveal the
non-local dependence of sub-system state functions on distant
interventions -- are suggested and discussed.
\end{abstract}

\maketitle

\section{Introduction}
\label{sec1}

The notion of ``weak measurement'', first introduced in
\cite{aav} and recently reviewed in \cite{boyd}, has 
become an important tool for
exploring foundational questions
in quantum mechanics.  For example, the recent theorem of Pusey,
Barrett, and Rudolph \cite{pbr} -- according to which the quantum state or wave
function must be understood as having an ontological (as opposed to
merely epistemic) character -- is nicely supported and supplemented by
the rather different recent work of Lundeen \emph{et al.}\ \cite{lundeen} showing that
the quantum wave function can be ``directly measured'' using weak
measurement techniques.
(This is ``direct'' in contrast to
the indirect or reconstructive approaches involved in quantum state
tomography -- but see also \cite{haap}). 

The procedure goes as follows.  In a weak measurement, one lets
a system in state $\ket{\psi}$ couple weakly to a pointer whose
position, if the coupling were stronger, would unambiguously register the value
associated with observable $\hat{A}$.  With the weak coupling, however,
the pointer's registration remains quite ambiguous; but this can be
made up for by repeating the process many times (on
identically-prepared systems) and averaging.  After
the system couples weakly to the pointer, one may also make a
(normal, strong) measurement of some other observable $\hat{B}$ and
post-select on the outcome.  In the weak-coupling limit, the average
value of the pointer's reading (when the final measurement has outcome
$b$) is the real part of the (here, complex) ``weak value''
\begin{equation}
\langle \hat{A} \rangle_W^b = \frac{ \bra{b} \hat{A} \ket{\psi}
}{\bracket{b}{\psi}}
\label{eq-wv}
\end{equation}
whose imaginary part is also accessible via measurements of the
pointer's conjugate momentum.  

In the scheme introduced by Lundeen \emph{et al.}, one lets $\hat{A} =
\hat{\pi}_x = \ket{x}\bra{x}$ and $\hat{B} = \hat{p}_x$.  We then have
that
\begin{equation}
\langle \hat{\pi}_x \rangle^{p_x}_W = \frac{ \bracket{p_x}{x} \bracket{x}{\psi}
}{\bracket{p_x}{\psi}} = \frac{e^{-i p_x x / \hbar} \psi(x)}{\tilde{\psi}(p_x)}.
\label{wf1}
\end{equation}
For the particular case $p_x = 0$ we thus have that the weak value is
proportional to the particle's wave function:
\begin{equation}
\langle \hat{\pi}_x \rangle_W^{p_x = 0} \sim \psi(x).
\end{equation}
Lundeen \emph{et al.}\ \cite{lundeen} used this technique to directly measure the transverse
wave function of a(n ensemble of identically prepared) photon(s).

Another recent example of the use of weak measurements to probe
foundational questions
involves Bohmian mechanics.  Wiseman \cite{wiseman} pointed out that a certain 
naively plausible operational approach to experimentally determining the trajectory of
a quantum particle -- namely, defining the velocity of a particle at a
certain position in terms of the difference between the weak value of
its position at time $t$ and the strong value at $t+dt$ -- yields
precisely the Bohmian expression for the particle's velocity (see also \cite{durr,oriols}). 
Steinberg \emph{et al.}\ \cite{steinberg} implemented this scheme to reconstruct the
average trajectories for photons in the 2-slit
experiment. 
The beautiful experimentally-reconstructed
trajectories are indeed congruent with the iconic images of
2-slit Bohmian trajectories \cite{dewdney}.  And it was recently
pointed out by Braverman and Simon \cite{braverman} that such 
measurements, if performed on one particle from
an entangled pair, should allow an empirical demonstration of the
non-local character of the Bohmian trajectories.

Following Braverman and Simon, the goal of the present work is to 
address the following seemingly
natural question:  what happens if the Lundeen \emph{et al.}\  technique,
for ``directly measuring'' the wave function of a particle, is applied
to a particle which does not, according to ordinary quantum theory,
\emph{have} a wave function of its own, because it is entangled with some other
particle(s)?  The answer turns out to be that, under suitable conditions, the ``directly
measured'' one-particle wave function corresponds exactly to the
so-called ``conditional wave function'' of Bohmian
mechanics \cite{dgz} . 
Since this is undoubtedly an unfamiliar concept to most
physicists, we review it in Section \ref{sec2} before explaining, in
Section \ref{sec3}, this central claim and suggesting an experimental
setup in which it should be demonstrable.  Section \ref{sec4} then
outlines a parallel result, regarding density matrices, especially appropriate
for the (spin) states of particles with discrete degrees of freedom.
Section \ref{sec5} offers conclusions, focusing especially on
questions surrounding the claimed observability of non-locality.

\section{Bohmian Conditional Wave Functions}
\label{sec2}

Consider for simplicity a system of two spin-0 particles (masses $m_1$ and
$m_2$, coordinates $x$ and
$y$) each moving in one spatial dimension.  According to ordinary
quantum mechanics (OQM) the wave function $\Psi(x,y,t)$, obeying an
appropriate two-particle Schr\"odinger equation, provides a
complete description of the state of the system.  According to Bohmian
mechanics (BM), however, the description provided by the wave function
alone is decidedly incomplete; a complete description requires
specifying in addition the actual particle positions $X(t)$ and
$Y(t)$.  For BM the wave function $\Psi(x,y,t)$ obeys the usual
Schr\"odinger equation, while $X(t)$  evolves
according to
\begin{equation}
\frac{d X(t)}{dt} = \left. \frac{\hbar}{2 m_1 i}\frac{ \Psi^*
  \frac{\partial}{\partial x} \Psi - \Psi \frac{\partial}{\partial x}
  \Psi^*}{\Psi^* \Psi} \right|_{x=X(t), \, y=Y(t)}
\label{dXdt}
\end{equation}
and similarly for $Y(t)$.  It is a joint property of the
time-evolution laws for the wave and particles that, if the particle
positions $X$ and $Y$ are random and $|\psi|^2$-distributed at some
initial time (this is the so-called quantum equilibrium hypothesis, QEH), 
they will remain $|\psi|^2$ distributed for all times.
This so-called ``equivariance'' property is crucial for understanding
how BM reproduces the statistical predictions of OQM \cite{dgz}.

The Bohmian ``conditional wave function'' (CWF) -- for, say, the 
first particle -- is simply the (``universal'') wave function
$\Psi(x,y,t)$ evaluated at $y = Y(t)$:
\begin{equation}
\chi_1(x,t) = \left. \Psi(x,y,t) \right|_{y=Y(t)}.
\end{equation}
This is the obvious and
natural way to construct a ``single particle wave function'' given the
resources that BM provides.  (OQM, with fewer resources at hand,
provides no such natural -- or even an unnatural -- construction.)  
What makes this definition natural is that the evolution law for the
position $X(t)$ of particle 1, Equation \eqref{dXdt},
 can be re-written in terms of particle 1's CWF as follows:
\begin{equation}
\frac{d X(t)}{dt} = \left. \frac{\hbar}{2 m_1 i} \frac{ \chi_1^*
  \frac{\partial}{\partial x} \chi_1 - \chi_1 \frac{\partial}{\partial
    x} \chi_1^* }{\chi_1^* \chi_1} \right|_{x=X(t)}.
\label{guidance2}
\end{equation}
It is thus appropriate to think of $\chi_1(x,t)$ as the guiding- or
pilot-wave that directly influences the motion of particle 1.

\begin{figure}[t]
\begin{center}
\scalebox{0.9}{
\scalebox{1} 
{
\begin{pspicture}(0,-4.33)(9.22,4.34)
\definecolor{color1329}{rgb}{0.6,0.6,0.6}
\definecolor{color1350g}{rgb}{0.0,0.4,1.0}
\definecolor{color1350f}{rgb}{0.0,0.8,1.0}
\definecolor{color1352g}{rgb}{1.0,1.0,0.2}
\definecolor{color1352f}{rgb}{1.0,1.0,0.6}
\psline[linewidth=0.02cm,linecolor=color1329](8.04,-3.36)(8.04,3.84)
\psline[linewidth=0.04cm,arrowsize=0.05291667cm 2.0,arrowlength=1.4,arrowinset=0.4]{<-}(1.64,4.24)(1.64,-3.76)
\usefont{T1}{ptm}{m}{n}
\rput(8.91,-3.655){$x$}
\usefont{T1}{ptm}{m}{n}
\rput(1.11,4.145){$y$}
\psline[linewidth=0.04cm](8.04,-3.16)(8.04,-3.56)
\usefont{T1}{ptm}{m}{n}
\rput(8.13,-3.855){$L$}
\usefont{T1}{ptm}{m}{n}
\rput(1.11,-1.935){$\lambda a_1$}
\usefont{T1}{ptm}{m}{n}
\rput(1.07,-0.135){$\lambda a_2$}
\usefont{T1}{ptm}{m}{n}
\rput(1.07,2.865){$\lambda a_3$}
\psellipse[linewidth=0.02,dimen=outer,fillstyle=gradient,gradlines=2000,gradbegin=color1350g,gradend=color1350f,gradmidpoint=1.0](4.84,-3.36)(3.2,0.2)
\psellipse[linewidth=0.02,dimen=outer,fillstyle=gradient,gradlines=2000,gradbegin=color1352g,gradend=color1352f,gradmidpoint=1.0](4.84,-1.96)(3.2,0.2)
\psellipse[linewidth=0.02,dimen=outer,fillstyle=gradient,gradlines=2000,gradbegin=color1352g,gradend=color1352f,gradmidpoint=1.0](4.84,-0.16)(3.2,0.2)
\psellipse[linewidth=0.02,dimen=outer,fillstyle=gradient,gradlines=2000,gradbegin=color1352g,gradend=color1352f,gradmidpoint=1.0](4.84,2.84)(3.2,0.2)
\psline[linewidth=0.04cm,arrowsize=0.05291667cm 2.0,arrowlength=1.4,arrowinset=0.4]{->}(1.24,-3.36)(8.84,-3.36)
\psdots[dotsize=0.12](5.94,-3.36)
\psdots[dotsize=0.12](5.5,-0.08)
\usefont{T1}{ptm}{m}{n}
\rput(3.93,1.345){$\{ X(0^+),Y(0^+) \}$}
\psbezier[linewidth=0.04,arrowsize=0.05291667cm 2.0,arrowlength=1.4,arrowinset=0.4]{->}(3.8,1.06)(4.2,0.28)(5.0,1.3)(5.38,0.14)
\usefont{T1}{ptm}{m}{n}
\rput(3.77,-2.535){$\{ X(0^-),Y(0^-) \}$}
\psline[linewidth=0.04cm](1.44,-1.96)(1.84,-1.96)
\psline[linewidth=0.04cm](1.44,-0.16)(1.84,-0.16)
\psline[linewidth=0.04cm](1.44,2.84)(1.84,2.84)
\psbezier[linewidth=0.04,arrowsize=0.05291667cm 2.0,arrowlength=1.4,arrowinset=0.4]{->}(5.1,-2.54)(5.54,-2.52)(5.54,-2.6)(5.84,-3.16)
\psline[linewidth=0.04cm,arrowsize=0.05291667cm 2.0,arrowlength=1.4,arrowinset=0.4]{<-}(8.04,0.84)(8.04,-0.56)
\psline[linewidth=0.04cm](1.64,-0.08)(8.04,-0.08)
\psbezier[linewidth=0.06](1.64,-0.1)(2.64,0.6)(3.8525,0.6)(4.84,-0.1)(5.8275,-0.8)(7.04,-0.8)(8.04,-0.1)
\usefont{T1}{ptm}{m}{n}
\rput(7.32,0.485){$\chi_1(x,0^+)$}
\psbezier[linewidth=0.06](1.68,-3.4)(2.2,-4.16)(2.34,-3.5)(2.82,-3.3)(3.3,-3.1)(3.3,-4.26)(3.58,-4.22)(3.86,-4.18)(3.58,-3.08)(4.06,-3.0)(4.54,-2.92)(4.9,-3.86)(6.02,-4.08)(7.14,-4.3)(6.66,-3.3)(7.04,-3.16)(7.42,-3.02)(7.28,-4.0)(8.04,-3.38)
\psline[linewidth=0.04cm,arrowsize=0.05291667cm 2.0,arrowlength=1.4,arrowinset=0.4]{<-}(8.04,-2.18)(8.04,-3.58)
\usefont{T1}{ptm}{m}{n}
\rput(7.26,-2.655){$\chi_1(x,0^-)$}
\end{pspicture} 
}
}
\caption{
Illustration of the collapse of the Bohmian CWF during an energy
measurement ($\hat{A} = \hat{H}$) on a particle in a box.  
The initial two-particle
wave function $\Psi(x,y,0^-) = \psi(x) \phi_0(y)$ has support in the
blueish region of the configuration space.  Since this initial state factorizes, the CWF at
$t=0^-$ is (up to a multiplicative constant) just $\chi_1(x) = \sum_n
c_n \psi_n(x)$, indicated
with the bolded lower curve.  The Schr\"odinger evolution from $0^-$
to $0^+$ produces a two-particle wave function with localized islands
of support in the configuration space, indicated by the yellowish
regions.  Each of the yellowish blobs is a Gaussian in $y$ (centered
at one of the possible post-interaction pointer positions $\lambda
a_n$) multiplied by one of the energy eigenfunctions (here
$\psi_{n}(x) \sim \sin(n \pi x / L)$).  And so if (for example, as
shown) the actual configuration point $\{ X(0^+), Y(0^+) \}$ ends up
in the support of the yellowish blob at $y=\lambda a_2$ -- something
which will occur with probability $|c_2|^2$ with random
initial configuration $\{ X(0^-),Y(0^-) \}$ in accord with the QEH -- then the
post-interaction CWF of particle 1 will be $\chi_1(x,0^+) \sim
\psi_{2}(x)$ (as shown).
\label{fig1}
}
\end{center}
\end{figure}
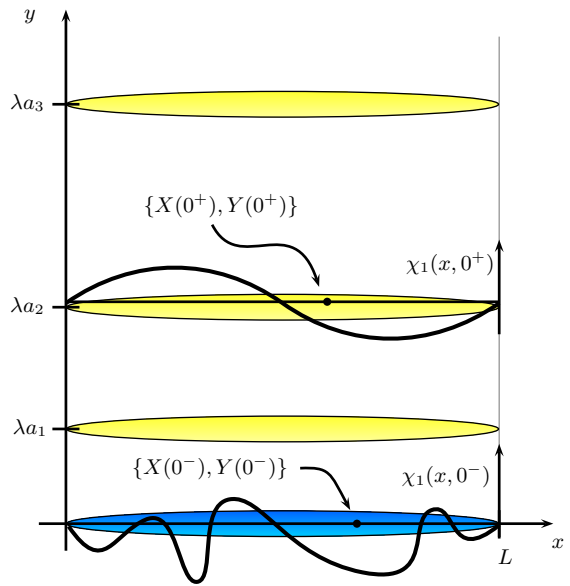

It is important to appreciate
that $\chi_1(x,t)$ depends on time in two different ways -- through
the $t$-dependence of $\Psi$ and also through the $t$-dependence of
$Y$.  Thus, in general, $\chi_1(x,t)$ does not obey a simple
one-particle Schr\"odinger equation, but obeys instead a more
complicated pseudo-Schr\"odinger equation \cite{dgz,telb}.   
In particular, it is easy to see that, under the appropriate measurement-like
circumstances, $\chi_1(x,t)$ will \emph{collapse}.  Suppose for
example that particle 1 has
initial wave function
\begin{equation}
\psi(x) = \sum_n c_n \psi_n(x)
\end{equation}
where the $\psi_n(x)$ are eigenstates of some observable $\hat{A}$
with eigenvalues
$a_n$.   And suppose that particle 2 is the pointer on an
$\hat{A}$-measuring device, initially in the state
\begin{equation}
\phi_0(y) \sim e^{-y^2/2w^2}.
\end{equation}
Now suppose the particles experience a (for simplicity, impulsive) interaction
\begin{equation}
\hat{H}_{int} =  \lambda \, \delta(t) \, \hat{A} \, \hat{p}_y. 
\end{equation}
The usual unitary Schr\"odinger evolution of the initial wave function
$\Psi(x,y,0^-) = \psi(x) \phi_0(y)$ then takes it into
\begin{equation}
\Psi(x,y,0^+) = \sum_n c_n \psi_n(x) \phi_0(y - \lambda a_n).
\label{wff}
\end{equation}
That is, the two-particle wave function after the interaction can be
understood as an entangled superposition of terms, each of which 
has particle 1 in an eigenstate of $\hat{A}$ and particle 2 in a new
position that registers the corresponding value $a_n$. 
(Note that we assume here
that $\lambda$ is sufficiently large that the separation between
adjacent values of $\lambda a_n$ is large compared to the width $w$ of
the pointer packet.  This is thus a ``strong'' measurement.)  
From the point
of view of OQM, Equation \eqref{wff} exhibits the standard problem of Schr\"odinger's cat:
instead of resolving the superposition of distinct $a$-values, the
measuring device itself gets infected with the superposition.  In OQM
(where there is nothing but the wave function at hand) one thus needs
to introduce additional dynamical (``collapse'') postulates to account
for the observed (apparently non-superposed) behavior of real
laboratory equipment.

In BM, however, there is no such problem.  The observable outcome of the
measurement is not to be found in the wave function, but instead in
the actual position $Y(0^+)$ of the pointer after the interaction.  It
is easy to see that (with appropriate random initial conditions) 
this will, with probability $|c_n|^2$, lie near the
value $\lambda a_n$ which indicates that the result of the measurement
was $a_n$.  Furthermore, it is easy to see that if $Y(0^+)$ is near the
value $\lambda a_n$, then the CWF of particle 1 will be (up to a
multiplicative constant) the appropriate eigenfunction:
\begin{equation}
\chi_1(x,0^+) \sim \psi_n(x).
\end{equation}
That is, the CWF of particle 1 \emph{collapses} (from a superposition
of several $\psi_n$s to the particular $\psi_n$ which corresponds to
the actually-realized outcome of the measurement) as a result of the
interaction with the measuring device, even though the
dynamics for the ``universal'' wave function $\Psi$ is completely
unitary.  See Figure \ref{fig1} and its caption for an illustration.

We have here explained the idea of (and one important and perhaps
surprising property of the dynamical evolution of) Bohmian CWFs as if
the particle of interest were interacting with the particle or
particles constituting a \emph{measuring device}.  That is of course the
crucial kind of situation if one is worrying about the so-called
quantum measurement problem.  But more important for our purposes here
is the fact that the Bohmian CWF (for a single particle) is perfectly
well-defined at all times for any particle that is part of a larger
(multi-particle) quantum system.  Indeed, BM only really provides a solution
of the measurement problem because it treats ``measurements'' as just
ordinary physical interactions, obeying the same universal dynamical laws as all
interactions.  It should thus be clear that, according to BM, collapses
(like the one we just described happening as a result of an interaction
with a measuring device) will actually be happening all the time, as
particles interact with each other.  It is the goal of the following
analysis to show how this feature of the Bohmian theory can be
experimentally manifested using weak measurement.

\section{Direct Measurement of Single Particle Wave Functions}
\label{sec3}

Let us then turn to the main result of the present paper.  Suppose
we carry out the Lundeen-type ``direct measurement of the wave function''
procedure on one particle of a two-particle system.    As a reality
check, suppose to begin with that the two-particle system has a
factorizable quantum state
\begin{equation}
\ket{\Psi} = \ket{\psi} \ket{\phi}
\end{equation}
where the first and second factors on the right refer to particles 1
and 2 respectively.  The Lundeen-type procedure involves
post-selecting on the final momentum $p_x$ of the particle whose wave
function we are trying to measure (here, particle 1).  Let us also post-select on the
final position $Y$ of particle 2 \cite{other}. 
It is then straightforward to calculate that
\begin{equation}
\langle \hat{\pi}_x \rangle_W^{p_x,y=Y} = \frac{ \bracket{Y}{\phi} \bracket{p_x}{x}
  \bracket{x}{\psi} }{\bracket{Y}{\phi} \bracket{p_x}{\psi}} =
\frac{e^{-i p_x x / \hbar} \, \psi(x)}{\tilde{\psi}(p_x)}
\end{equation}
which is, as expected, identical to Equation \eqref{wf1}.  

If, however, particle 1 is in a general, entangled state with particle 2, as in
\begin{equation}
\ket{\Psi} = \int dx' dy' \Psi(x',y') \ket{x'} \ket{y'}
\end{equation}
then the operational
determination of particle 1's wave function (post-selected on the
final strongly-measured position $Y$ of particle 2) yields
\begin{eqnarray}
\langle \hat{\pi}_x \rangle^{p_x,y=Y}_W &=& \frac{ \bracket{p_x}{x}
  \bracket{x,Y}{\Psi}}{ \bracket{p_x,Y}{\Psi}} \nonumber \\ 
&=& \frac{e^{-i p_x x / \hbar} \, \Psi(x,Y)}{\int dx' \Psi(x',Y)\, e^{-i
    p_x x' / \hbar} } 
\end{eqnarray}
Restricting, as before, our attention to the cases in which the final
measured momentum $p_x$ is zero, we have that
\begin{equation}
\langle \hat{\pi}_x \rangle_W^{p_x = 0, y=Y} \sim \Psi(x,Y) = \chi_1(x)
\end{equation}
where the right hand side is precisely the Bohmian CWF for particle
1.  Note that we have tacitly relied on the fact that, for Bohmian
mechanics, position is a \emph{non-contextual} (``hidden'') variable.  Thus
the final position measurement on particle 2 simply
reveals, for Bohmian mechanics, the actual pre-existing location
$Y$ of that particle.  In short, the two $Y$s in the analysis -- the one
representing the outcome of the final position measurement of particle
2, and the one, used in the definition of the Bohmian CWF,
representing the actual position of particle 2 -- are, for Bohmian
mechanics, the same.

So far we have basically ignored the issue of the exact timing
of the various measurements on the two-particle system.  Let us then
examine this in the context of a somewhat
concrete example.  Consider two
photons prepared in some kind of entangled state (to be specified
shortly) and propagating in roughly opposite directions.  Let the variable
$x$ refer to the transverse spatial degree of freedom of photon 1
(propagating, say, to the right) and the variable $y$ refer to the
transverse spatial degree of freedom of photon 2 (propagating to the
left).  We imagine a setup like that reported in \cite{lundeen} in
which the weak measurement is effected using an extremely narrow
half-wave plate (``$\lambda/2$ sliver'') and the $p_x = 0$ 
post-selection is effected by accepting only those photons which pass
a narrow slit downstream from and along the axis of a Fourier Transform lens.  The
remaining photons are then passed through an appropriate quarter (or
half) wave plate; the imbalance between the two polarization states
then yields the real (or, respectively, imaginary) part of the
transverse wave function at the location of the $\lambda/2$ sliver.  See \cite{lundeen} for details.

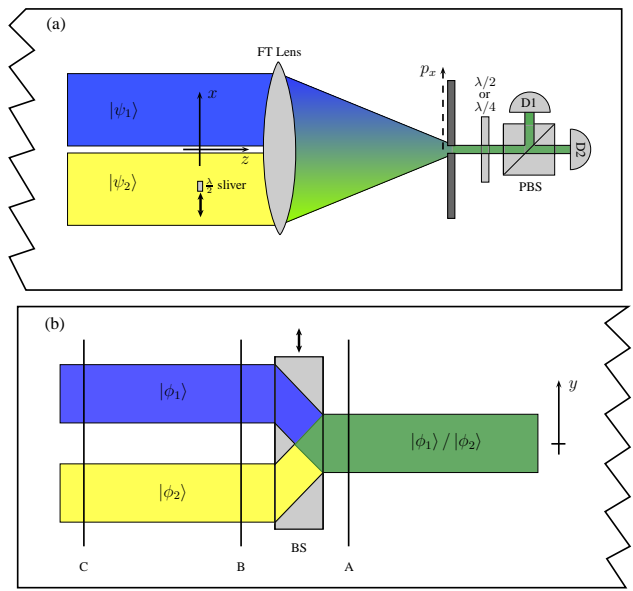
\begin{figure}[t]
\begin{center}
\scalebox{0.55}{
\scalebox{1} 
{
\begin{pspicture}(0,-7.02)(15.02,7.02)
\definecolor{color111b}{rgb}{1.0,1.0,0.3333333333333333}
\definecolor{color112b}{rgb}{0.396078431372549,0.6666666666666666,0.396078431372549}
\definecolor{color113b}{rgb}{0.23137254901960785,0.3333333333333333,1.0}
\definecolor{color114b}{rgb}{1.0,1.0,0.32941176470588235}
\definecolor{color115g}{rgb}{0.1803921568627451,0.22745098039215686,1.0}
\definecolor{color115f}{rgb}{0.6078431372549019,1.0,0.011764705882352941}
\definecolor{color116b}{rgb}{0.8,0.8,0.8}
\definecolor{color138b}{rgb}{0.39215686274509803,0.6666666666666666,0.39215686274509803}
\definecolor{color138}{rgb}{0.6,0.6,0.6}
\definecolor{color144b}{rgb}{0.3333333333333333,0.6666666666666666,0.3333333333333333}
\definecolor{color152b}{rgb}{0.4,0.4,0.4}
\definecolor{color167b}{rgb}{0.32941176470588235,0.3411764705882353,1.0}
\definecolor{color168b}{rgb}{0.30196078431372547,0.3137254901960784,1.0}
\psframe[linewidth=0.01,dimen=outer,fillstyle=solid,fillcolor=color111b](6.42,-4.0)(1.22,-5.42)
\psframe[linewidth=0.01,dimen=outer,fillstyle=solid,fillcolor=color112b](12.78,-2.8)(7.58,-4.22)
\psframe[linewidth=0.01,dimen=outer,fillstyle=solid,fillcolor=color113b](6.6,5.44)(1.4,3.68)
\psframe[linewidth=0.01,dimen=outer,fillstyle=solid,fillcolor=color114b](6.6,3.52)(1.4,1.76)
\pspolygon[linewidth=0.01,fillstyle=gradient,gradlines=2000,gradbegin=color115g,gradend=color115f,gradmidpoint=1.0](6.6,5.4)(6.6,1.8)(11.0,3.6)
\psbezier[linewidth=0.02,fillstyle=solid,fillcolor=color116b](6.5,5.72)(6.54,5.72)(6.9021516,5.0398407)(6.92,3.62)(6.9378486,2.2001593)(6.58,1.54)(6.54,1.54)(6.5,1.54)(6.14,2.22)(6.14,3.62)(6.14,5.02)(6.46,5.72)(6.5,5.72)
\usefont{T1}{ptm}{m}{n}
\rput(6.53,5.965){$\text{FT Lens}$}
\psline[linewidth=0.04cm,linestyle=dashed,dash=0.16cm 0.16cm,arrowsize=0.05291667cm 2.0,arrowlength=1.4,arrowinset=0.4]{->}(10.48,3.6)(10.48,5.6)
\usefont{T1}{ptm}{m}{n}
\rput(10.13,5.49){\large $p_x$}
\usefont{T1}{ptm}{m}{n}
\rput(11.5,5.225){$\lambda/2$}
\usefont{T1}{ptm}{m}{n}
\rput(11.49,4.925){$\text{or}$}
\usefont{T1}{ptm}{m}{n}
\rput(11.5,4.605){$\lambda/4$}
\usefont{T1}{ptm}{m}{n}
\rput(12.6,2.705){$\text{PBS}$}
\pscircle[linewidth=0.02,dimen=outer,fillstyle=solid,fillcolor=color116b](12.58,4.5){0.5}
\psframe[linewidth=0.01,linecolor=white,dimen=outer,fillstyle=solid](13.16,4.5)(11.94,3.94)
\rput{-90.0}(9.96,17.16){\pscircle[linewidth=0.02,dimen=outer,fillstyle=solid,fillcolor=color116b](13.56,3.6){0.5}}
\rput{-90.0}(9.65,16.91){\psframe[linewidth=0.01,linecolor=white,dimen=outer,fillstyle=solid](13.89,3.91)(12.67,3.35)}
\psframe[linewidth=0.0060,linecolor=color138,dimen=outer,fillstyle=solid,fillcolor=color138b](13.56,3.7)(10.7,3.5)
\psframe[linewidth=0.02,dimen=outer,fillstyle=solid,fillcolor=color116b](11.6,4.4)(11.4,2.8)
\psframe[linewidth=0.02,dimen=outer,fillstyle=solid,fillcolor=color116b](13.19,4.23)(11.93,2.97)
\psframe[linewidth=0.0040,linecolor=color138,dimen=outer,fillstyle=solid,fillcolor=color138b](11.58,3.7)(11.42,3.5)
\psframe[linewidth=0.0060,linecolor=color138,dimen=outer,fillstyle=solid,fillcolor=color138b](13.2,3.7)(11.96,3.5)
\pspolygon[linewidth=0.0060,linecolor=color138,fillstyle=solid,fillcolor=color138b](12.46,3.52)(12.68,3.71125)(12.68,4.2)(12.46,4.2)
\psframe[linewidth=0.0060,linecolor=color138,dimen=outer,fillstyle=solid,fillcolor=color144b](12.68,4.5)(12.46,4.18)
\psline[linewidth=0.02cm](12.08,4.5)(13.08,4.5)
\psline[linewidth=0.02cm](13.56,4.1)(13.56,3.1)
\psline[linewidth=0.01cm](10.74,3.7)(12.46,3.7)
\psline[linewidth=0.01cm](10.76,3.5)(13.56,3.5)
\psline[linewidth=0.01cm](12.46,4.5)(12.46,3.7)
\psline[linewidth=0.01cm](12.68,4.5)(12.68,3.7)
\psline[linewidth=0.01cm](12.7,3.7)(13.56,3.7)
\psframe[linewidth=0.025999999,dimen=outer,fillstyle=solid,fillcolor=color152b](10.78,5.28)(10.58,3.68)
\psframe[linewidth=0.025999999,dimen=outer,fillstyle=solid,fillcolor=color152b](10.78,3.52)(10.58,1.92)
\usefont{T1}{ptm}{m}{n}
\rput(12.56,4.745){$\text{D1}$}
\usefont{T1}{ptm}{m}{n}
\rput{-90.0}(10.21,17.39){\rput(13.8,3.605){$\text{D2}$}}
\usefont{T1}{ptm}{m}{n}
\rput(2.77,4.55){\large $\ket{\psi_1}$}
\usefont{T1}{ptm}{m}{n}
\rput(2.75,2.79){\large $\ket{\psi_2}$}
\pspolygon[linewidth=0.04](0.4,7.0)(14.8,7.0)(14.8,0.2)(0.4,0.2)(0.0,1.0)(0.6,1.2)(0.0,2.2)(0.6,2.4)(0.0,3.4)(0.6,3.6)(0.0,4.6)(0.6,4.8)(0.0,5.6)(0.6,6.0)(0.0,6.6)
\pspolygon[linewidth=0.04](14.96,-0.2)(0.2,-0.2)(0.2,-7.0)(14.9,-7.0)(14.44,-6.3)(15.0,-6.0)(14.4,-5.16)(15.0,-4.8)(14.4,-4.0)(15.0,-3.6)(14.4,-2.8)(15.0,-2.4)(14.4,-1.6)(15.0,-1.2)(14.4,-0.6)
\usefont{T1}{ptm}{m}{n}
\rput(10.56,-3.47){\large $\ket{\phi_1} \, / \, \ket{\phi_2}$}
\psframe[linewidth=0.03,dimen=outer,fillstyle=solid,fillcolor=color116b](7.6,-1.4)(6.4,-5.6)
\psframe[linewidth=0.01,dimen=outer,fillstyle=solid,fillcolor=color167b](6.42,-1.6)(1.22,-3.02)
\pspolygon[linewidth=0.01,linecolor=color138,fillstyle=solid,fillcolor=color168b](6.42,-1.6)(7.58,-2.82)(7.58,-4.22)(6.42,-3.02)
\pspolygon[linewidth=0.01,linecolor=color138,fillstyle=solid,fillcolor=color111b](6.42,-5.44)(7.58,-4.22)(7.58,-2.84)(6.42,-4.02)
\psline[linewidth=0.03cm](6.42,-1.4)(6.42,-5.6)
\psline[linewidth=0.01cm](6.42,-1.6)(7.6,-2.84)
\psline[linewidth=0.01cm](6.42,-5.42)(7.6,-4.2)
\psline[linewidth=0.01cm](6.4,-3.0)(6.9,-3.52)
\psline[linewidth=0.01cm](6.9,-3.52)(6.4,-4.04)
\usefont{T1}{ptm}{m}{n}
\rput(3.97,-2.27){\large $\ket{\phi_1}$}
\usefont{T1}{ptm}{m}{n}
\rput(3.95,-4.69){\large $\ket{\phi_2}$}
\usefont{T1}{ptm}{m}{n}
\rput(7.0,-6.035){$\text{BS}$}
\usefont{T1}{ptm}{m}{n}
\rput(1.13,6.61){\large $\text{(a)}$}
\usefont{T1}{ptm}{m}{n}
\rput(1.1,-0.65){\large $\text{(b)}$}
\psline[linewidth=0.04cm,arrowsize=0.05291667cm 2.0,arrowlength=1.4,arrowinset=0.4]{<-}(4.6,5.0)(4.6,3.2)
\psline[linewidth=0.04cm,arrowsize=0.05291667cm 2.0,arrowlength=1.4,arrowinset=0.4]{->}(4.2,3.6)(5.8,3.6)
\usefont{T1}{ptm}{m}{n}
\rput(5.7,3.33){\large $z$}
\usefont{T1}{ptm}{m}{n}
\rput(4.91,4.93){\large $x$}
\psframe[linewidth=0.02,dimen=outer,fillstyle=solid,fillcolor=color116b](4.68,2.84)(4.54,2.58)
\usefont{T1}{ptm}{m}{n}
\rput(5.22,2.725){\small $\frac{\lambda}{2} \text{ sliver}$}
\psline[linewidth=0.04cm](8.2,-1.0)(8.2,-6.0)
\psline[linewidth=0.04cm](1.8,-1.0)(1.8,-6.0)
\psline[linewidth=0.04cm](5.6,-1.0)(5.6,-6.0)
\usefont{T1}{ptm}{m}{n}
\rput(8.21,-6.495){$\text{A}$}
\usefont{T1}{ptm}{m}{n}
\rput(5.6,-6.495){$\text{B}$}
\usefont{T1}{ptm}{m}{n}
\rput(1.8,-6.495){$\text{C}$}
\pspolygon[linewidth=0.0020,linecolor=OliveGreen,fillstyle=solid,fillcolor=color138b](7.58,-2.82)(6.9,-3.54)(7.58,-4.22)
\psline[linewidth=0.02cm](11.94,4.2)(11.94,3.0)
\psline[linewidth=0.06cm,arrowsize=0.053cm 1.6,arrowlength=1.2,arrowinset=0.25]{<->}(4.62,2.56)(4.62,1.96)
\psline[linewidth=0.06cm,arrowsize=0.053cm 1.6,arrowlength=1.2,arrowinset=0.25]{<->}(7.0,-0.72)(7.0,-1.32)
\psline[linewidth=0.03cm](7.58,-1.42)(7.58,-5.58)
\psline[linewidth=0.02cm](11.96,3.0)(13.16,4.2)
\psline[linewidth=0.02cm](13.18,3.7)(13.18,3.48)
\psline[linewidth=0.04cm,arrowsize=0.05291667cm 2.0,arrowlength=1.4,arrowinset=0.4]{<-}(13.28,-1.98)(13.28,-3.78)
\psline[linewidth=0.04cm](13.1,-3.52)(13.44,-3.52)
\usefont{T1}{ptm}{m}{n}
\rput(13.63,-2.13){\large $y$}
\end{pspicture} 
}
}
\caption{
Schematic diagram of the proposed experiment.  Frame (a), following
\cite{lundeen}, shows photon 1 propagating to the right and
undergoing first a weak measurement (effected by the $\lambda/2$
sliver), then a $p_x=0$ measurement/post-selection, and finally the
readout of the weak measurement.  Frame (b) shows photon 2 propagating
to the left.  The two-photon state is $\ket{\Psi} = \frac{1}{\sqrt{2}}
\left( \ket{\psi_1} \ket{\phi_1} + \ket{\psi_2} \ket{\phi_2} \right)$
where $\ket{\psi_1}$ and $\ket{\psi_2}$ have distinct (transverse,
i.e., $x$) spatial profiles -- for example, as shown here, perhaps
$\ket{\psi_1}$ has support only for $x>0$ while $\ket{\psi_2}$ has
support only for $x<0$.  For photon 2, $\ket{\phi_1}$ and
$\ket{\phi_2}$ are initially overlapping spatially but are distinct in
some property (such as energy or polarization) which allows the beams
to be separated, by some kind of (removable) beam-splitter (BS), as shown.  Three different possible detection planes
-- A, B, and C -- for the final measurement/post-selection of the
transverse position $Y$ of photon 2 are shown and discussed in the main text.
\label{fig2}
}
\end{center}
\end{figure}

We now consider the possibility that each photon that enters the device
is entangled with a second photon:
\begin{equation}
\ket{\Psi} = \frac{1}{\sqrt{2}} \left( \ket{\psi_1} \ket{\phi_1} +
  \ket{\psi_2} \ket{\phi_2} \right).
\label{state}
\end{equation}
It is important (for the proper functioning of the wave function
measurement) that $\ket{\psi_1}$ and $\ket{\psi_2}$ have the same
(say, linear) polarization.  But let them have distinct (transverse)
spatial profiles -- for example, and most simply, suppose that
$\ket{\psi_1}$ has transverse spatial support just \emph{above} the
$z$-axis (i.e., for $x>0$) while $\ket{\psi_2}$ has transverse spatial support just
\emph{below} the $z$-axis ($x<0$).  See Figure \ref{fig2}(a).

As to photon 2, suppose that (at least initially) $\ket{\phi_1}$ and
$\ket{\phi_2}$ have identical transverse profiles and 
completely overlap spatially, but are distinct in some
way (for example, they could be orthogonally polarized, or could have
different energies) that allows the two parts of the beam to be
separated by some type of beam splitter (BS).   See Figure
\ref{fig2}(b).  

Let us now consider several different spatial locations for, and
time-orderings involving, the final measurement of the position $Y$ of
photon 2.  

To begin with, let us first imagine that the
measurement/post-selection on photon 2's transverse coordinate $y$
occurs (in, say, the lab frame) \emph{before} the weak measurement on
photon 1 and at a plane like A in Figure \ref{fig2} (i.e., before
photon 2 has passed any beam splitter).  According to
OQM, this measurement of $Y$ will collapse the 2-particle wave
function and leave photon 1 with a definite (non-entangled) wave
function of its own.  Because $\ket{\phi_1}$ and $\ket{\phi_2}$
overlap at A, however, the position measurement gives no information
about $\ket{\phi_1}$ vs. $\ket{\phi_2}$ and so leaves photon 1 in the
state $\frac{1}{\sqrt{2}} \left( \ket{\psi_1} + \ket{\psi_2}
\right)$.   And this of course coincides with the predicted result of
the direct measurement protocol:  
for measurement/post-selection at A
\begin{eqnarray}
\langle \hat{\pi}_x \rangle_W^{p_x = 0,y=Y} &=& \frac{ \psi_1(x)
  \bracket{Y}{\phi_1} + \psi_2(x)
  \bracket{Y}{\phi_2}}{\tilde{\psi}_1(0) \bracket{Y}{\phi_1} +
  \tilde{\psi}_2(0) \bracket{Y}{\phi_2} } \nonumber \\
&\sim& \frac{1}{\sqrt{2}} \left( \psi_1(x) + \psi_2(x) \right)
\end{eqnarray}
where we have used the fact that, for detection at the A plane,
$\bracket{Y}{\phi_1} = \bracket{Y}{\phi_2}$.

On the other hand, if the measurement/post-selection on
photon 2 is performed at plane B (after photon 2 has passed the beam
splitter, but still \emph{before} the measurement protocol
has been carried out on photon 1) then according to OQM the wave
function of photon 1 will collapse to \emph{either} $\ket{\psi_1}$ (if
$Y$ is found in the support of $\bracket{y}{\phi_1}$) \emph{or}
$\ket{\psi_2}$ (if $Y$ is found in the support of $\bracket{y}{\phi_2}$).
This again coincides with the expected results of the direct measurement:
for detection at the B plane, one or the other of
$\bracket{Y}{\phi_1}$ and $\bracket{Y}{\phi_2}$ will be zero.  We will
thus find that
\begin{equation}
\langle \hat{\pi}_x \rangle_W^{p_x = 0, y=Y} \sim \psi_m(x)
\label{collapse}
\end{equation}
for $Y \in \text{supp}(\bracket{y}{\phi_m})$.

For both of those two scenarios, ordinary QM attributes a one-particle
wave function to particle 1 at the time in question; this wave
function coincides with the Bohmian CWF and the expected results of
the weak measurement technique.  In short, there is nothing surprising
or interesting here from the point of view of OQM.

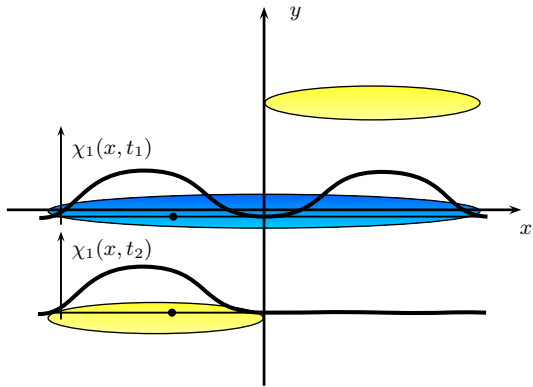
\begin{figure}[t]
\begin{center}
\scalebox{0.9}{
\scalebox{1} 
{
\begin{pspicture}(0,-2.85)(7.98,2.87)
\definecolor{color270g}{rgb}{0.0,0.4,1.0}
\definecolor{color270f}{rgb}{0.0,0.8,1.0}
\definecolor{color272g}{rgb}{1.0,1.0,0.2}
\definecolor{color272f}{rgb}{1.0,1.0,0.6}
\usefont{T1}{ptm}{m}{n}
\rput(7.67,-0.525){$x$}
\usefont{T1}{ptm}{m}{n}
\rput(4.27,2.675){$y$}
\psellipse[linewidth=0.02,dimen=outer,fillstyle=gradient,gradlines=2000,gradbegin=color270g,gradend=color270f,gradmidpoint=1.0](3.8,-0.25)(3.2,0.26)
\psellipse[linewidth=0.02,dimen=outer,fillstyle=gradient,gradlines=2000,gradbegin=color272g,gradend=color272f,gradmidpoint=1.0](2.2,-1.83)(1.6,0.24)
\psellipse[linewidth=0.02,dimen=outer,fillstyle=gradient,gradlines=2000,gradbegin=color272g,gradend=color272f,gradmidpoint=1.0](5.4,1.35)(1.6,0.26)
\psline[linewidth=0.04cm,arrowsize=0.05291667cm 2.0,arrowlength=1.4,arrowinset=0.4]{->}(0.0,-0.23)(7.6,-0.23)
\psdots[dotsize=0.12](2.46,-0.33)
\psdots[dotsize=0.12](2.44,-1.75)
\psline[linewidth=0.027999999cm](0.6,-1.75)(7.0,-1.75)
\psline[linewidth=0.027999999cm,arrowsize=0.05291667cm 2.0,arrowlength=1.4,arrowinset=0.4]{<-}(0.8,-0.55)(0.8,-1.85)
\usefont{T1}{ptm}{m}{n}
\rput(1.56,-0.825){$\chi_1(x,t_2)$}
\psline[linewidth=0.027999999cm,arrowsize=0.05291667cm 2.0,arrowlength=1.4,arrowinset=0.4]{<-}(0.8,1.01)(0.8,-0.45)
\usefont{T1}{ptm}{m}{n}
\rput(1.56,0.675){$\chi_1(x,t_1)$}
\psline[linewidth=0.027999999cm](0.68,-0.33)(7.08,-0.33)
\psbezier[linewidth=0.06](0.48,-0.35)(1.0539885,-0.37)(1.0731214,0.37)(2.06,0.35)(3.0468786,0.33)(2.8333533,-0.32871726)(3.809133,-0.33)(4.7849126,-0.33128273)(4.5935836,0.35)(5.58,0.33)(6.5664163,0.31)(6.4494796,-0.31)(7.1,-0.33)
\psbezier[linewidth=0.06](0.46,-1.77)(1.0339885,-1.79)(1.0531213,-1.05)(2.04,-1.07)(3.0268786,-1.09)(2.8133533,-1.7487173)(3.7891328,-1.75)(4.7649126,-1.7512827)(4.5735836,-1.73)(5.58,-1.75)(6.5864162,-1.77)(6.4294796,-1.73)(7.08,-1.75)
\psline[linewidth=0.04cm,arrowsize=0.05291667cm 2.0,arrowlength=1.4,arrowinset=0.4]{<-}(3.8,2.77)(3.8,-2.83)
\end{pspicture} 
}
}
\caption{
In the $(x,y)$ configuration space, the entangled two-photon wave
function initially has support in the blueish region and the CWF for
photon 1 -- $\chi_1(x,t_1)$ -- looks something like the two-hump
curve shown.   The passage of photon 2 through the beam-splitter (BS)
separates the two-photon wave function into the two yellowish
islands.  The actual configuration point (shown here as a black dot)
ends up (depending on its random initial position for each photon
pair) in one of the two yellowish islands, and the CWF at
$t_2$ will thus have collapsed to either $\bracket{x}{\psi_1}$ or
$\bracket{x}{\psi_2}$.  (The latter case is shown here.)  
\label{fig3}
}
\end{center}
\end{figure}

Consider, however, a scenario
involving measurement/post-selection
of photon 2's position at plane C (such that this measurement takes
place \emph{well after} the weak measurement on photon 1 has already
gone fully to completion).  It is trivial to see that the results of
the weak measurement will again be given by Equation \eqref{collapse}
-- that is, dramatically different (``collapsed'') wave functions for
photon 1 will be found depending on whether particle 2 is (later!) found
in the support of $\ket{\phi_1}$ or $\ket{\phi_2}$.  From the point of
view of OQM, however, it is rather difficult to understand why, prior
to any actual measurement (meaning here an interaction involving
macroscopic amplification) that would trigger a collapse, one should
find collapsed one-particle wave functions.  On the other hand, this
is perfectly natural from a Bohmian point of view:  as sketched in
Figure \ref{fig3} the conditional wave function (CWF) for particle 1
collapses as soon as the packets separate in the two-dimensional
configuration space.  This happens when the components $\ket{\phi_1}$
and $\ket{\phi_2}$ are split at the BS; no actual position
\emph{measurement} is required.  (Note that the claim here is
\emph{not} that OQM makes the wrong predictions.  Undoubtedly it makes
the right predictions -- indeed we have used nothing but orthodox
quantum ideas to calculate what should be observed in the experiment.
The point is rather only that, from the OQM perspective, it is at best
obscure what one ought to expect the operationalist determination of
photon 1's ``one-particle wave function'' to yield at times when
photon 1 remains entangled with photon 2.  Whereas, from the Bohmian
point of view, the results are the obvious, natural thing.)  

The proposed experiment, then, should involve a fixed
detection plane, like plane C in the Figure, at a greater optical
distance from the two-particle source than that of the
measuring apparatus for particle 1.  The removable beam splitter BS
should, on the other hand, ideally be slightly closer to the
two-particle source than the particle 1 apparatus.
With the BS in place, the arrangement would be
like that shown in the Figure and the ``direct measurement'' on photon
1 would yield (after appropriate post-selection on $Y$) collapsed
photon 1 wave functions (even though the actual position measurement
on photon 2 would occur well after photon 1 was already measured).  On the other
hand, with the BS removed, the situation would be equivalent to
detecting photon 2 at plane A (except that this too would only occur
later, after the measurements on photon 1 had occured) and the
measurement of photon 1 would reveal the uncollapsed wave function.
One could thus in some sense observe the collapse of the Bohmian CWF of photon 1 
as a direct result of the insertion (perhaps at space-like separation)
of the BS into the path of photon 2.  

This proposed  setup should be realizable in practice along
the following lines.  Type II spontaneous parametric down-conversion yields a pair
of photons in an entangled polarization state $\left( \ket{H} \ket{H}
  + \ket{V} \ket{V} \right) / \sqrt{2}$ with the individual photons being
coupled into single-mode optical fibres.  The first photon
should then be split by a polarizing beam splitter (PBS), with, say, the $\ket{H}$ component
being shunted into the $+z$ direction with $x > 0$ and the $\ket{V}$
component passed through a $\lambda/2$ plate (rotating its
polarization to $\ket{H}$) before being shunted into the $+z$
direction with $x<0$.  This prepares photon 1 as suggested in Figure
\ref{fig2}(a) and the subsequent measurements may then be carried out
exactly as in  \cite{lundeen}.  The second photon can be directly shunted into
the $-z$ direction so that the two orthogonal polarization components
have identical and perfectly overlapping transverse spatial profiles,
as in Figure \ref{fig2}(b).  The overall (transverse spatial and polarization)
two-photon state after such preparation can thus be written
\begin{equation}
\ket{\Psi} = \frac{1}{\sqrt{2}} \left( \ket{+H} \ket{\emptyset H} +
  \ket{-H} \ket{\emptyset V} \right)
\label{eq20}
\end{equation}
where $+$ indicates that the transverse state has support for $x>0$,
$-$ indicates that the support lies in $x<0$, and $\emptyset$
indicates that the support is centered at $x=0$.  (It is also of
course understood that photon 1 is moving in the $+z$ direction
and photon 2 in the $-z$ direction.)  The perfect correspondence between
Equations \eqref{eq20} and \eqref{state} should be clear.
Note that for this type of implementation, the
(generic) ``BS'' in Figure \ref{fig2}(b) can be a standard PBS.

\section{Density Matrices}
\label{sec4}

In the Lundeen \emph{et al.}\ procedure for measuring the wave
function, it is the particles' spatial degree of freedom that is
probed, with the polarization serving as the pointer.  But the
polarization state of a particle can also be measured using weak
measurement techniques, with the position degree of freedom (or in
principle some other extrinsic degree of freedom) playing the role of
the pointer.  For example, in a scheme recently proposed and
demonstrated by Lundeen and Bamber \cite{lb,bl} (see also
\cite{boyd2}) the polarization density matrix
of a particle can be measured as follows.  For a photon in a mixed
state described by density
operator $\hat{\rho}$, the weak value of an observable $\hat{A}$ is
given by 
\begin{equation}
\langle \hat{A} \rangle^b_W = \frac{ \langle b | \hat{A} \hat{\rho} | b
  \rangle}{\langle b | \hat{\rho} | b \rangle}
\label{eq-wvmixed}
\end{equation}
where $\ket{b}$ is the final post-selected state, as in Equation
\eqref{eq-wv}.  In the case of a pure state, $\hat{\rho} =
\ket{\psi}\bra{\psi}$, Equation \eqref{eq-wvmixed} reduces to
\eqref{eq-wv}, whereas for a genuinely mixed state, \eqref{eq-wvmixed}
is the appropriate weighted average.  In the event of no
post-selection, a further averaging gives
\begin{equation}
\langle \hat{A} \rangle_W = \text{Tr} \left[ \hat{A} \hat{\rho} \right].
\label{wv-mixed}
\end{equation}
The Lundeen/Bamber procedure can then be most simply understood
as follows.  Defining operators
\begin{equation}
\hat{\pi}_{ij} = \ket{i} \bra{j}
\label{eq-pi}
\end{equation}
(with $i,j \in \{ H, V \}$)  
on the two-dimensional polarization Hilbert space for a photon, one
sees that their weak values correspond to the entries in the
polarization density matrix:
\begin{equation}
\langle \hat{\pi}_{ij} \rangle_W = \text{Tr}\left[ \hat{\pi}_{ij} \hat{\rho}
\right] = \bra{j} \hat{\rho} \ket{i}.
\end{equation}
Of course, for $i \ne j$, $\hat{\pi}_{ij}$ is not a Hermitian
operator, so measuring it -- even weakly -- raises some questions.
But the worrisome matrix elements can be re-expressed in terms of weak
values of perfectly reputable operators by introducing post-selection, e.g.,
\begin{eqnarray}
\rho_{\, V \! H} &=& \bra{H} \hat{\rho} \ket{V} \nonumber \\
&=& P(D) \, \langle \hat{\pi}_{HH}\rangle_W^{D} - P(A) \, \langle \hat{\pi}_{HH}
\rangle_W^{A}
\end{eqnarray}
where $\ket{D} = (\ket{H}+\ket{V})/\sqrt{2}$ and $\ket{A} = (\ket{H} -
\ket{V})/\sqrt{2}$, and $P(D) = \langle D | \hat{\rho} | D \rangle$ and $P(A) =\langle A | \hat{\rho} | A \rangle$ are respectively the rates
of successful post-selection on the $\ket{D}$ and $\ket{A}$ states.  
See \cite{lb} for further details.

Consider now a two-particle system in state
\begin{equation}
\ket{\psi_o} = \int dy \sum_{i,j} \psi_{i,j}(y)  \ket{i}_1 \ket{j}_2 \ket{y}_2
\end{equation}
where, as before, $i,j \in \{ H, V\}$ are one-particle polarization
eigenstates, and $\ket{y}_2$ is a position eigenstate of particle 2.
(We suppress, for simplicity, the position degree of freedom of
particle 1; recall that it may be used as the pointer variable 
to weakly measure the polarization density matrix.)
The two-particle state thus has density operator
\begin{equation}
\hat{\rho} = \ket{\psi_0} \bra{\psi_0}
\end{equation}
in terms of which one can define the reduced density matrix (RDM) of
particle 1 by tracing over the degrees of freedom associated with
particle 2:
\begin{equation}
\hat{\rho}_1^{\,\text{red}} = \int d y \; \text{Tr}_2 \left[
  \bra{y} \hat{\rho} \ket{y} \right] =
\int dy \sum_{j} \, \bra{y, j} \hat{\rho} \ket{y,j}.
\end{equation}
The RDM is of course the standard way of defining
the ``state'' of a (perhaps-entangled) subsystem.  

As discussed above, for systems with complex-valued wave functions, Bohmian
mechanics allows one to define the conditional wave function of a
sub-system, in terms of which the guidance law for the particles
comprising the sub-system can be re-expressed.  For systems with
discrete (spin, polarization) degrees of freedom, however, the Bohmian
CWF for each one-particle sub-system would carry the discrete indices for
\emph{all} particles in the system.  It thus cannot really be regarded
as a ``wave function for a single particle''.  In such situations,
Bohmian mechanics thus follows ordinary quantum mechanics in defining
the state of the sub-system in terms of an appropriate density
matrix.  But as was first pointed out by Bell \cite{bell}, the correct
Bohmian particle trajectories (needed to reproduce the statistical
predictions of ordinary QM) cannot be expressed in terms of the usual
reduced density matrix.  Instead, one needs to introduce the
``conditional density matrix'' (CDM), which involves tracing over the
discrete indices associated with particles outside the sub-system in
question, but then evaluating the spatial variables (again, associated
with particles outside the sub-system in question) at the actual
locations of the Bohmian particles.  (See \cite{dm} for a
detailed discussion.) Thus, for the system introduced
just above, we would have
\begin{equation}
\hat{\rho}^{\, \text{cond}}_1 = \text{Tr}_2 \left[ \bra{Y} \hat{\rho}
\ket{Y} \right].
\label{bcdm}
\end{equation}
Note that a sub-system will in general possess both a RDM and a CDM,
but that these will not in general be equal:  the RDM can be
understood as the \emph{average} of all possible CDMs. (If we had normalized the CDM, then the RDM would be given by the average of all possible CDMs, weighted by the usual quantum probability for $y=Y$.)

Now, what should happen if one performs the Lundeen/Bamber procedure
for directly measuring the polarization density matrix of one photon
from an entangled pair, also -- as in the previous section --
post-selecting on the final position $Y$ of the second particle?  It
is easy to see that under such conditions the weak value of an
operator $\hat{A}$ (acting just on the particle 1 Hilbert space) is
\begin{eqnarray}
\langle \hat{A} \rangle^Y_W &=& \text{Tr}_{1,2} \! \left[ \bra{Y} \, \hat{A} \,
  \hat{\rho} \, \ket{Y} \right] \nonumber \\
&=& \text{Tr}_1 \left[ \hat{A} \; \text{Tr}_2 \! \left[ \bra{Y}\,  \hat{\rho}
    \, \ket{Y} \right] \right] \nonumber \\
&=& \text{Tr}_1 \! \left[ \hat{A} \, \hat{\rho}^{\text{cond}}_1 \right].
\end{eqnarray}
The last line is identical to Equation \eqref{wv-mixed}
except that the (one-particle) density matrix $\hat{\rho}$ is replaced
by the Bohmian CDM from Equation \eqref{bcdm}.  One thus expects that
the operational procedure sketched above -- in which $\hat{A} =
\hat{\pi}_{ij}$ from Equation \eqref{eq-pi} -- should yield the
Bohmian conditional density matrix (and not the usual reduced density
matrix) as the directly-measured one-particle density matrix.

Of course, just as with the setup discussed in the previous
section, this result is not very interesting or surprising if the
post-selection-basing measurement of particle 2's position $Y$ occurs
prior to the weak measurement procedure on particle 1.  For then, the
measurement of $Y$ will have collapsed the two-particle state such
that the ordinary RDM and the Bohmian CDM coincide.  

On the other hand, if one arranges for the measurement of $Y$ to occur
only \emph{after} the procedure on particle 1 has gone to completion,
it is quite interesting indeed that the procedure should yield the
Bohmian CDM as opposed to the ordinary RDM.  Consider for example the
following setup, very much in the spirit of the one proposed in the
previous section.  A two-photon system is prepared in the state
\begin{equation}
\ket{\psi_1} = \ket{\phi^0} \left( \ket{H}_1 \ket{H}_2 + \ket{V}_1
  \ket{V}_2 \right)/\sqrt{2}
\end{equation}
where $\ket{\phi^0}$, referring to the transverse spatial degree of
freedom of particle 2, is (say) a Gaussian centered at $y=0$.  (The
spatial degrees of freedom of particle 1, and the non-transverse
spatial degrees of freedom of particle 2, are suppressed for
simplicity.)  By means of a polarizing beam splitter that can be
inserted (or not) in the path of particle 2, the two-particle state
may (or may not) be transformed into
\begin{equation}
\ket{\psi_2} = \left( \ket{\phi^+} \ket{H}_1 \ket{H}_2 +
  \ket{\phi^-} \ket{V}_1 \ket{V}_2 \right) / \sqrt{2} 
\end{equation}
where $\ket{\phi^\pm}$ is (say) a Gaussian displaced in the $\pm$
y-direction by an amount that is larger than its width.  

The crucial point is then that, for both $\ket{\psi_1}$ and
$\ket{\psi_2}$, the RDM for particle 1 is
\begin{equation}
\hat{\rho}^{\, \text{red}}_1 \xrightarrow[\ket{H},\ket{V} \,
\text{basis}]{} \left( \begin{array}{cc} 
\sfrac{1}{2} & 0 \\
0 & \sfrac{1}{2} \end{array} \right).
\end{equation}
The Bohmian CDM for particle 1 will be (proportional to) this same
matrix if the state is $\ket{\psi_1}$.  But if the beam splitter
is inserted such that the state is $\ket{\psi_2}$, the CDM will have
``collapsed'', being now proportional to either 
\begin{equation}
\hat{\rho}^{\, \text{cond}}_1 \xrightarrow[\ket{H},\ket{V} \,
\text{basis}]{} \left( \begin{array}{cc} 1 & 0 \\ 0 & 0 \end{array}
\right) \;  \text{OR}  \; \left( \begin{array}{cc} 0 & 0 \\ 0 &
    1 \end{array} \right)
\end{equation}
depending on whether $Y \in \text{supp}(\bracket{y}{\phi^+})$ or $Y \in
\text{supp}(\bracket{y}{\phi^-})$.  

According to Bohmian mechanics, one of
these two possibilities is realized -- and the particle 1 CDM
collapses accordingly -- as soon as particle 2 traverses
the polarizing beam splitter (should it be inserted).  Furthermore,
once the $\ket{H}$ and $\ket{V}$ components of the particle 2 beam are
split apart, the actual particle position $Y$ will not, according to
the theory, change.  So the actual measurement of $Y$ (for the purpose
of post-selection) can wait as long as is desired -- for example,
until after the weak measurement procedure on particle 1 has been
carried out.  Still, when the dust settles and all the data is
properly binned up, OQM predicts that it is a collapsed density matrix
corresponding precisely to the Bohmian CDM, that should be revealed by
the direct measurement of particle 1's state.

\section{Discussion}
\label{sec5}

If the procedure of
 \cite{lundeen} for making a ``direct
measurement of the quantum wave function'' 
is applied to one particle from an entangled
pair (and regarded as a plausible operationalist definition of the
``single particle wave function'' for such a particle) the result,
with suitable post-selection on the other particle, is
precisely the ``conditional wave function'' (CWF) of Bohmian mechanics --
that is, the natural theoretical concept of a ``single particle wave
function'' that Bohmian mechanics (uniquely) makes possible.
Similarly, the results of applying a related procedure -- for directly measuring the density
matrix associated with a single particle -- to one particle from an
entangled two-particle system, should yield the Bohmian ``conditional
density matrix'' (CDM) as opposed to the more standard reduced density
matrix (RDM).  These results are
particularly interesting when the weak measurement (that reveals the state of
particle 1) is carried out prior to the strong position measurement on
particle 2 on which post-selection will be based.  Thus, in the same
way that Braverman and Simon \cite{braverman} have suggested that one can ``observe the
nonlocality of Bohmian trajectories with entangled photons''
 one should also be able to observe the nonlocal
dependence  of Bohmian single-particle states (wave functions and
density matrices) on distant interventions such as the insertion (or
not) of the BS in Figure \ref{fig2}.

Indeed, from the
perspective suggested earlier, in which each particle's CWF (or CDM)  is
regarded as the object which directly guides or pilots the Bohmian particle,
the present work can be seen as digging yet one level
deeper beyond Braverman/Simon: 
instead of merely observing how the particle trajectories
change as a result of some distant interventions, one may also observe
how the ``field'' \emph{responsible} for those changes itself
changes.  A successful experimental demonstration of this effect would
thus in some sense reveal the non-local character of Bohmian mechanics
in an unprecedentedly fundamental way. 

But to formulate things in these ways is to invite several possible
misunderstandings, so let us clarify a couple of points.  

First, as Einstein \cite{einstein} famously remarked: ``Whether you can
observe a thing or not depends on the theory which you use.  It is the
theory which decides what can be observed.''   In particular, the
claims above should only be understood to mean that \emph{from the point of
view of Bohmian mechanics}, the experimental procedures outlined here
can be understood as \emph{observations} of (for example) the
interesting dynamical behavior of single particle CWFs.  From the
point of view of orthodox QM, on the other hand, the same procedures
would evidently \emph{not} be regarded as genuine observations of
anything.  The situation is thus completely parallel to that
surrounding the reconstructed particle trajectories in
\cite{steinberg}.  The result of the empirical procedure 
coincides with certain things that are posited to actually exist by
one candidate theory, Bohmian mechanics.  But another candidate
theory, which does not posit those things, also has no trouble
accounting for the results.  Indeed, insofar as the two theories are
empirically equivalent, the results cannot really count as evidence
for or against either theory.  Nevertheless, the expected results
remain somehow more natural from the Bohmian point of view!
The situation can perhaps be summed up
like this:  \emph{if} one accepts, from the outset, that
empirically measured weak values correspond to some physically real
features of the system in question, then the predicted results would
be trivial to reconcile with Bohmian mechanics but difficult to
reconcile with ordinary QM.  But of course, whether one should accept
such a premise is, to put it mildly, highly controversial.  

And then, second:  It would be easy to get the impression from the way
things were put above that
by inserting (or refraining from inserting) the BS in the path of
photon 2, one can instantaneously affect the observable CWF/CDM of the
distant particle 1.  That is, it would be easy to get the impression
that one could (in principle, if impractically) send a superluminal
signal by running many copies of the experiment in parallel (so that
the many trials required to build up sufficient data all occur
simultaneously).   But
this, of course, should be impossible (whether one believes in BM 
or OQM or any other such empirically-equivalent theory).  

There are two points to be understood here, one rather obvious and one
more subtle.  The obvious point is that Alice (on the right) must
learn the outcome of Bob's position measurement (on the left) before
she can know how to properly bin her data.  And this information will
have to be sent to her through a ``classical'' (i.e., here,
sub-luminal) communication channel.  So it is already clear that no
actual superluminal signalling will be possible.  

The more subtle, and more interesting, point is that the statistical
relationship between Alice's and Bob's measurement results is actually
independent of the exact temporal sequence of the measurements.  This
is certainly not surprising from the point of view of relativity,
given that Alice's and Bob's measurements may well occur at spacelike
separation.  But it is somewhat surprising from the point of view of
Bohmian mechanics, which involves a hidden (but dynamically relevant)
privileged reference frame.  

Essentially for reasons of drama, we have described the setup above,
in the allegedly interesting cases, so
that the temporal sequence is as follows:  first Bob (or his
assistant) decides whether to insert or not insert the BS into the
path of particle 2; then
Alice's measurement protocol on particle 1 occurs; and then finally Bob measures the
final transverse position $Y$ of particle 2.  From the point of view
of Bohmian mechanics, then, we may say the following.  If this is the true
temporal sequence -- in the dynamically privileged reference frame
posited by the theory -- then things develop causally in the way we
have suggested:  the insertion of the BS (if and only if it is
inserted) causes the two-particle wave function to divide in the
configuration space as shown in
Figure \ref{fig3} and thus causes the CWF (or CDM) for photon 1 to collapse;
the (here, subsequent) measurement protocol by Alice then simply 
reveals the true CWF/CDM of photon 1 at the time of that measurement; the
final measurement/post-selection by Bob then plays the (dynamically) purely
passive role of revealing, to Bob, what was already physically
definite, in order that the already-acquired data can be properly
binned.

But since the privileged frame is, for Bohmian mechanics, hidden, it
is entirely possible that the ``true'' temporal sequence (i.e., the
temporal sequence in the privileged frame) is instead as follows:
Alice's measurement protocol on photon 1 occurs \emph{first};
\emph{then} comes Bob's (assistant's) decision to insert the BS or
not, followed by measurement of the transverse position $Y$.  If this
is the ``true'' temporal sequence, the \emph{statistics} will be
unchanged, but the causal story will be somewhat different.  To
wit:  instead of the passage (or not) of photon 2 through the BS
affecting the CWF/CDM of photon 1, now it will be the measurement protocol
on photon 1 which (at least sometimes) affects the CWF/CDM of photon 2 \emph{and thus
  influences where, for a given $Y$, it will go, should it encounter
  the BS}.  Concretely, there
will exist possible initial conditions for the 2-particle system 
which have the following property:  had the measurement protocol on
particle 1 not been carried out, particle 2 would definitely have gone
``up'' at the BS (and would hence have been found with $Y>0$), but given that the measurement protocol on particle 1
\emph{was} carried out, particle 2 instead went ``down'' at the BS
(and was hence found with $Y<0$).  

With this second possible ``true'' temporal sequence, it is no longer
really the case that the weak measurement protocol on particle 1 is
simply revealing the structure of particle 1's CWF at the time of the
measurement.  Instead, the measurement on particle 1 may actively affect
the state (in particular, the CWF or CDM) 
of particle 2, making the subsequent post-selection on
particle 2's position rather less benign, less passive.  And this
makes clear in principle why, 
despite the presence in the theory of a
dynamically privileged reference frame in which instantaneous
action-at-a-distance occurs, one is not only prevented from
sending signals faster than light, but also 
prevented from putting any experimental limits on the speed of the
laboratory with respect to the presumed underlying privileged
reference frame.  The statistical patterns in the data will --
presumably -- remain the same as the ``true'' temporal sequence is
varied between the two possibilities discussed here, even as the
Bohmian causal story changes rather dramatically.

It does, however, remain absolutely valid to say that -- provided one
adopts the Bohmian point of view -- the experimental
setup suggested here would allow for a direct empirical observation of
the non-local dependence of a single particle's (Bohmian) CWF/CDM on
distant interventions.  It's just that -- compared to the way we
initially explained things -- it is rather ambiguous whether one is
observing the effect, on particle 1's CWF/CDM, of inserting or not
inserting the BS in front of particle 2 ... or instead observing the effect, on particle 2's
CWF/CDM, of carrying out the weak measurement protocol on 
particle 1.  It thus remains appropriate
to conclude that a realization of the proposed type of experiment would
be quite interesting and would in particular help bring the Bohmian 
CWF/CDM into the light of experimental reality.  

\begin{acknowledgments}
W.S.\ acknowledges support of a grant from the John Templeton Foundation. The opinions expressed in this publication are those of the authors and do not necessarily reflect the views of the John Templeton Foundation.
\end{acknowledgments}

\end{document}